\documentclass[aoas,nameyear,dvips]{arximspdf}
\usepackage{accents}
\usepackage{graphicx}
\usepackage{breakurl}

\doi{10.1214/10-AOAS353}
\volume{4}
\issue{4}
\pubyear{2010}
\firstpage{1774}
\lastpage{1796}

\makeatletter
\newcommand{\Nbold}{\mathbf{N}_{\mathrm{obs}}}
\newcommand{\Nrowbold}{\mathbf{N}_{\mathrm{row}}}
\newcommand{\Ncolbold}{\mathbf{N}_{\mathrm{col}}}
\newcommand{\Ni}{\mathrm{N_i}}
\newcommand{\Nimiss}{\bolds{N}_{\mathrm{miss}_{\mathrm {i}}}}
\newcommand{\Mimiss}{\bolds{M}_{\mathrm{miss}_{\mathrm{i}}}}
\newcommand{\Nicomp}{\bolds{N}_{\mathrm{comp}_{\mathrm{i}}}}
\newcommand{\uNirow}{{\underaccent{\sim}{\mathrm{N}}}{}_{\mathrm{row}_\mathrm{i}}}
\newcommand{\uNicol}{{\underaccent{\sim}{\mathrm{N}}}{}_{\mathrm{col}_\mathrm{i}}}
\newcommand{\Nirc}{N_{{\mathrm{rc}}_{\mathrm{i}}}}
\newcommand{\Mirc}{M_{{\mathrm{rc}}_{\mathrm{i}}}}
\newcommand{\Nib}{{\mathrm{N_{b_i}}}}
\newcommand{\Niw}{{\mathrm{N_{w_i}}}}
\newcommand{\Nih}{{\mathrm{N_{h_i}}}}
\newcommand{\NibD}{N_{\mathrm{bD_i}}}
\newcommand{\NibR}{N_{\mathrm{bR_i}}}
\newcommand{\NiwD}{N_{\mathrm{wD_i}}}
\newcommand{\NiwR}{N_{\mathrm{wR_i}}}
\newcommand{\NihD}{N_{\mathrm{hD_i}}}
\newcommand{\NihR}{N_{\mathrm{hR_i}}}
\newcommand{\NibA}{N_{\mathrm{bA_i}}}
\newcommand{\NiwA}{N_{\mathrm{wA_i}}}
\newcommand{\NihA}{N_{\mathrm{hA_i}}}
\newcommand{\NiD}{{\mathrm{N}_{\mathrm{D_i}}}}
\newcommand{\NiR}{{\mathrm{N}_{\mathrm{R_i}}}}
\newcommand{\NiA}{{\mathrm{N}_{\mathrm{A_i}}}}
\newcommand{\BibD}{\beta_{\mathrm{bD_i}}}
\newcommand{\uBi}{{\underaccent{\sim}{\mathrm{\beta}}}{}_\mathrm{i}}
\newcommand{\Xih}{\mathrm{X}_{\mathrm{h_i}}}
\newcommand{\uthi}{{\underaccent{\sim}{\mathrm{\theta}}}{}_{\mathrm{i}}}
\newcommand{\uthri} {{\underaccent{\sim}{\theta}}{}_{\mathrm{r}_\mathrm{i}}}
\newcommand{\uomi}{{\underaccent{\sim}{\omega}}{}_{\mathrm{i}}}
\newcommand{\thrci}{\theta_{\mathrm{rc}_{\mathrm{i}}}}
\newcommand{\umu}{{\underaccent{\sim}{\mu}}}
\newcommand{\umuz}{{\underaccent{\sim}{\mu}}{}_{0}}
\newcommand{\SI}{\bolds{\Sigma}}
\newcommand{\SIb}{\bolds{\Sigma}_{\mathrm{b}}}
\newcommand{\SIbw}{\bolds{\Sigma}_{\mathrm{bw}}}
\newcommand{\SIw}{\bolds{\Sigma}_{\mathrm{w}}}
\newcommand{\SIh}{\bolds{\Sigma}_{\mathrm{h}}}
\newcommand{\SIbh}{\bolds{\Sigma}_{\mathrm{bh}}}
\newcommand{\SIwh}{\bolds{\Sigma}_{\mathrm{wh}}}
\newcommand{\SIba}{\bolds{\Sigma}_{\mathrm{ba}}}
\newcommand{\SIwa}{\bolds{\Sigma}_{\mathrm{wa}}}
\newcommand{\SIha}{\bolds{\Sigma}_{\mathrm{ha}}}
\newcommand{\SIa}{\bolds{\Sigma}_{\mathrm{a}}}
\newcommand{\PSz}{\bolds{\Psi}_0}
\newcommand{\nuz}{\nu_0}
\newcommand{\kazbold}{\bolds{\kappa}_0}
\newcommand{\LArc}{\Lambda_{\mathrm{rc}}}
\newcommand{\LAhD}{\Lambda_{\mathrm{hD}}}
\newcommand{\TOrc}{\mathrm{TO}_{\mathrm{rc}}}
\newcommand{\GAr}{\Gamma_{\mathrm{r}}}
\newcommand{\Ki}{\mathrm{K_i}}
\newcommand{\uKi}{\mathrm{{\underaccent{\sim}{K}}_i}}
\newcommand{\Kirc}{\mathrm{K}_{\mathrm{rc_i}}}
\newcommand{\Kibold}{\mathbf{K}_{\mathrm{i}}}
\newcommand{\Kbold}{\mathbf K}
\newcommand{\Ji}{\mathrm{J_i}}

\makeatother

\begin{document}
\begin{frontmatter}

\title{Exit polling and racial bloc voting: Combining individual-level
and R${}\times{}$C ecological data}
\runtitle{Exit polling and racial bloc voting}

\begin{aug}
\author[A]{\fnms{D. James} \snm{Greiner}\ead[label=e1]{jgreiner@law.harvard.edu}\corref{}}
\and
\author[B]{\fnms{Kevin M.} \snm{Quinn}\ead[label=e2]{kquinn@law.berkeley.edu}}

\runauthor{D. J. Greiner and K. M. Quinn}
\affiliation{Harvard Law School and UC Berkeley School of Law}
\address[A]{Harvard Law School\\
504 Griswold Hall\\
Cambridge, Massachusetts 02138\\
USA\\
\printead{e1}} 
\address[B]{UC Berkeley School of Law\\
490 Simon Hall\\
Berkeley, California 94720-7200\\
USA\\
\printead{e2}}
\end{aug}

\received{\smonth{6} \syear{2009}}
\revised{\smonth{4} \syear{2010}}

%
\begin{abstract}
Despite its shortcomings, cross-level or ecological inference
remains a necessary part of some areas of quantitative inference,
including in United States voting rights litigation. Ecological
inference suffers from a lack of identification that, most agree, is
best addressed by incorporating individual-level data into the
model. In this paper we test the limits of such an incorporation
by attempting it in the context of drawing inferences about racial
voting patterns using a combination of an exit poll and
precinct-level ecological data; accurate information about racial
voting patterns is needed to assess triggers in voting rights laws
that can determine the composition of United States legislative
bodies. Specifically, we extend and study a hybrid model that
addresses two-way tables of arbitrary dimension. We apply the
hybrid model to an exit poll we administered in the City of Boston
in 2008. Using the resulting data as well as simulation, we compare
the performance of a pure ecological estimator, pure survey
estimators using various sampling schemes and our hybrid. We
conclude that the hybrid estimator offers substantial benefits by
enabling substantive inferences about voting patterns not
practicably available without its use.
\end{abstract}

%
\begin{keyword}
\kwd{Ecological inference}
\kwd{Bayesian inference}
\kwd{voting rights litigation}
\kwd{exit polls}
\kwd{survey sampling}.
\end{keyword}

\end{frontmatter}

Cross-level or ecological inference is the attempt to draw conclusions
about statistical relationships at one level from data aggregated to a
higher level. Frequently, ecological inference is conceptualized as
the attempt to infer individual-level relationships from a set of
contingency tables when only the row and column totals are observed.
One important application of ecological inference is in United States
redistricting litigation, in which a critical issue is whether the
voting patterns of racial groups differ. Because the secret ballot
prevents direct observation of voter races and voter choices,
redistricting litigants and their experts are ordinarily required to
attempt to infer racial voting patterns by examining election returns
(reported at the precinct or perhaps the ``vote tabulation district''
level) as married to demographic information from the Decennial Census.
In this paper we explore issues associated with incorporating
individual level information, in the form of responses to an exit poll
we administered in the City of Boston, into an R${}\times{}$C ecological
model.

Speaking broadly, the lack of identification in ecological models was
famously discussed in \citet{Robinson1950}. Since then, most to
consider the question have agreed that, if ecological inference is to
be attempted, the best way to proceed is to incorporate additional,
preferably individual-level (from a survey), information into the
model. In the sampling literature, combining survey results with
population-level information has a long and rich history, dating back
at least to \citet{DemingStephan1940}.
\citet{BishopFienbergHolland1975} (see pages 97--102) discuss what they
call the ``classical'' use of iterative proportional fitting, also
called ``raking'' to known marginal totals or ``incomplete
post-stratification'' [see \citet{DevilleEtAl1993}]. Additional
examples include \citet{BelinEtAl1993}, \citet{Little1993} and
\citet{Zaslavsky1993}. As relevant to this paper, the idea is that two
categorical measurements are made on each of K in-sample units so that
estimated counts are generated for each of the cells of an R${}\times{}$C
two-way contingency table. These estimated internal cell counts are
adjusted so as to conform to row and column sums known from another
source.
The quirk in ecological data is that there are many such
contingency tables (precincts, in our application) and, as to some of
those, one row category so dominates the table
count as to make bounding information [see \citet{DuncanDavis1953}]
informative, rendering sampling in the precincts that correspond to these tables a waste of resources.
(Were the column sums
known in advance, the same principle might apply, but in our
application the column sums are vote totals that are unknown in
advance.) We explore such issues of sample allocation in this paper.

Meanwhile, on the social science side, the past decade or so has seen
several papers [\citet{SteelEtAl2003}, \citet{RaghunathanEtAl2003},
\citet{GlynnEtAl2008}, \citet{HaneuseWakefield2008},
\citet{GlynnEtAl2009}] addressing how best to combine ecological data
with limited individual-level information. As is true in ecological
inference more generally, most papers addressing incorporation of
additional information into ecological data have focused on sets of $2
\times2$ contingency tables, which (after conditioning on the row and
column totals) involve one missing quantity per table.

In this paper we address the R${}\times{}$C case,\footnote{Software to
implement the methods we propose, including those used in this paper,
is available via the R package ``RxCEcolInf''; access CRAN from
\url{http://www.r-project.org/}.} building on earlier work of our own
[\citet{GreinerQuinn2009}], which in turn built on
\citet{BrownPayne1986} and \citet{Wakefield2004}. We do so because of
the importance of R${}\times{}$C ecological inference to many fields of inquiry;
a particular interest of ours is in United States voting rights
litigation. Sections 2 and 5 of the Voting Rights Act prevent dilution
and retrogression of the voting strength of racial (``racial'' means
racial or ethnic) minorities via gerrymandering of districts. In both
settings, proof that members of different racial groups vote similarly
within-group and differently between-group constitutes part of what is
called ``racially polarized'' or ``racial bloc'' voting, which is the
``keystone'' to litigation [\citet{11thCir1984}] and the ``undisputed
and unchallenged center'' [\citet{Issacharoff1992}] to the area of law.
This law can in turn decide the composition of Congress as well as of
local legislative bodies [\citet{Lublin1995}]. Thus, the need for
accurate information regarding the voting preferences of different
racial groups is acute. As mentioned above, because the secret ballot
prevents direct observation of voter decisions, proof of racial bloc
voting is most frequently made via R${}\times{}$C ecological inference
methods, as follows. The Census provides voting-age-population figures
for each racial group, which are arranged along the rows of contingency
tables. Table columns are official vote counts for each candidate
(Democrat, Republican, etc.), along with an additional ``Abstain''
column to account for persons declining to exercise the franchise.
There is one such contingency table for each voting precinct, and the
goal of the inference is to calculate, say, the percentage of Hispanic
voters who voted for the Democrat (see below for more technical
definitions). That requires filling in the missing internal cell
counts of the contingency tables, subject to the constraints imposed by
the row and column totals.

The need for better and more accurate techniques in this area has grown
in recent years. As the number of relevant racial and ethnic groups in
the United States polity increases (from, say, black versus white to
include Hispanics and Asians), inference becomes more complicated.
Additional races represent additional rows in the contingency tables,
requiring more parameters in a model and imposing greater challenges at
the model-fitting stage [\citet{Greiner2006}]. Incorporating
individual-level information from a survey into the R${}\times{}$C
ecological inference model represents one promising avenue in this
area.

We accordingly subject the task of combining individual-level and R
$\times$ C ecological data to a stress test in the form of an effort to
draw inferences about the voting behavior of R racial groups using data
aggregated to the level of the precinct together with an exit poll in
which not all precincts were in-sample. Specifically, we discuss the
challenges, choices and results of a 400-pollster, 11-university,
39-polling-place exit poll we administered in the City of Boston on the
November~4, 2008 election. Combining ecological data with an exit poll
constitutes a stress test for a hybrid model because (i) the nature of
exit polling prevents us from implementing optimal subsampling
techniques recently explored in the literature, (ii)~survey nonresponse
is ever-present, and (iii) the fact that several precincts may be
combined within a single voting location requires additional
assumptions regarding the aggregation process, as we explain below. In
our view, our hybrid model passes this stress test by supporting
substantive conclusions, particularly regarding voting behavior of
hard-to-estimate groups such as Asian- and Hispanic-Americans, that
could not be reached without its use (all of this assuming the
reasonableness of the model).

We organize this paper as follows: we clarify notation before
presenting a brief taxonomy of R${}\times{}$C ecological techniques that
focuses on the advantages and disadvantages of fraction versus count
models. We articulate the details of our hybrid ecological/survey
proposal and use simulation to study its behavior, focusing in
particular on its performance in the presence of aggregation bias,
defined immediately below. On the basis of these simulations, we offer
guidance for practitioners confronted with a choice of three classes of
estimators: an ecological model alone, a survey sample alone, and a
hybrid. We demonstrate that (i) the hybrid is always preferable to the
ecological model; (ii) in the absence of severe aggregation bias, the
hybrid dominates the survey sample estimator; (iii) in the presence of
severe aggregation bias, the hybrid is still probably preferable,
although the researcher's choice of estimator depends on, among other
things, whether the contingency tables tend to be dominated by one row
(in voting applications, this corresponds to a high level of housing
segregation), and whether interest lies primarily in the point estimate
or valid intervals.

We then present the process leading to and the results of our City of
Boston exit poll, focusing on voting behavior by race in a
Massachusetts ballot initiative regarding marijuana (other results
from this exit poll are available from the authors on request). We
demonstrate that our hybrid estimator allows inferences unavailable
from either the exit poll or the ecological inference model alone.
Without the hybrid estimator, for example, little can be said regarding
Asian-American voting preferences in Boston, nor can one easily
distinguish between Hispanic and white preferences. We also find
little evidence of aggregation bias in the Boston data.

Regarding the definition of aggregation bias, the critical assumption
of most ecological inference techniques is the absence of contextual
effects. Contextual effects can occur when the distribution of the
internal cell counts varies with the distribution of the allocation of
the counts by row. In voting parlance, if white voting behavior varies
with the fraction of whites in the precinct, this contextual effect
will cause the aggregation process to induce bias in almost any
ecological estimator, unless a covariate/predictor can be included in
the model to remove this effect.

Regarding notation, any quantity with the subscript $_{\mathrm{rc}_{\mathrm{i}}}$
refers to that quantity in the ith contingency table's (precinct's) rth
row, cth column. In our application, r can be b for black, w for
white, h for Hispanic or a for Asian; c can be D for Democrat, R for
Republican or A for Abstain (meaning choosing not to vote).
\textit{N}'s, \textit{M}'s and K's refer to counts, as follows: \textit{N}'s
are the unobserved, true internal cell counts; K's are the counts as
observed in the survey; and $\Mirc= \Nirc- \Kirc$. We italicize
unobserved counts but leave observed quantities in ordinary typescript.
Table \ref{tbl:3.3.counts.full} clarifies our representations for the
case of $3 \times3$ precinct tables involving African-American,
Caucasian and Hispanic groups in a Democrat versus Republican contest.

\begin{table}
\caption{3 $\times$ 3 table of voting by race}\label{tbl:3.3.counts.full}
\begin{tabular}{@{}lcccc@{}}
\hline
& \textbf{Dem} & \textbf{Rep} & \textbf{Abstain} & \\
\hline
Black & $\NibD$ & $\NibR$ & $\NibA$ & $\Nib$\\
White & $\NiwD$ & $\NiwR$ & $\NiwA$ &$\Niw$\\
Hispanic & $\NihD$ & $\NihR$ & $\NihA$ & $\Nih$ \\[5pt]
& $\NiD$ & $\NiR$ & $\NiA$& $\textrm{N}_{\mathrm{i}}$\\
\hline
\end{tabular}
\end{table}

We further suppose that a survey or exit poll is implemented in a
subset S of the I precincts in the jurisdiction and contest of
interest. In precinct i $\in$ S, $\Kibold$ is a random matrix of
dimension $\Ji\,\times$ (R${}\times{}$C), where $\Ji$ is the number of individuals
surveyed in this precinct. Each row of $\Kibold$ is a vector of 0's
except for a 1 corresponding to the cell of the precinct contingency
table in which the surveyed individual belongs, where the cells are
vectorized row major. In the Table \ref{tbl:3.3.counts.full} example,
a vector $(0, 0, 0, 0, 0, 1, 0, 0, 0)$ would indicate a white person
who abstained from voting. Let $\Kbold$ represent a matrix of all of
the $\Kibold$'s (organized in any coherent manner).

Let $\uNirow$ ($\uNicol$) represent the vector of observed row
(column) totals in the ith precinct, with $\Nrowbold$ ($\Ncolbold$) a
matrix of all $\uNirow$'s ($\uNicol$'s), and $\Nbold=
[{{\Nrowbold\enskip \Ncolbold}}]$. Let $\Nicomp$ equal the (unobserved)
full set of internal cell counts in the ith precinct. Finally, let
$\Nimiss$ denote any set of (R$-1)\times{}$(C$-1$) counts for the ith precinct
which, had they been observed in conjunction with $\uNirow$ and
$\uNicol$, would have been sufficient to determine all table counts.
In Table \ref{tbl:3.3.counts.full}, for example, $\Nimiss$ could
equal $\bigl[{{ \NibD\enskip \NibR}\atop {\NiwD\enskip \NiwR}}\bigr]$.
Note that $\Nicomp$ and $\Nimiss$ are used in the missing data sense
[e.g., \citet{LittleRubin2002}].

Finally, because our interest is primarily in ecological inference as
opposed to survey methods, we do not investigate potential biases in
surveys or exit polls, except to compare the predictions of our City of
Boston exit poll to the observed results. That comparison suggests an
encouraging absence of systematic biases, including the absence of a
``Bradley'' effect for Obama versus McCain, a result in accord with
recent findings [\citet{Hopkins2008}]. Moreover, while we acknowledge
the potential for a variety of sources of bias in ecological studies
[see \citet{SalwayWakefield2005} for a review], we focus our attention
on aggregation bias, which we believe to be potentially most
problematic in this area [\citet{Rivers1998}].

\section{Fraction versus count models}\label{s1}

We discuss briefly some advantages and disadvantages of modeling
unobserved internal cell counts as opposed to the fractions produced
when a researcher divides these unobserved counts by their
corresponding row totals.

Apart from the approach we advocate, a variety of R${}\times{}$C
ecological models have been proposed: for example, the unconstrained
[see \citet{AchenShively1995}] or constrained [\citet{Gelman2001}]
linear model, the truncated multivariate normal proposal in
\citet{King1997}, the Dirichlet-based method in \citet{Rosen2001}, and
the information theoretic proposal in \citet{Judge2004}. These other
proposals all share the feature that they model (at various levels) not
the internal cell counts themselves, but rather the fractions produced
when the unobserved internal cell counts are divided by their row
totals. In contrast, we model internal cell counts. There are
strengths and weaknesses to each approach.

Formally, let $\beta$'s refer to the (unobserved) internal cell
fractions so $\BibD= \frac{\NibD}{\Nib}$, and $\uBi$ refer to the
vector of the $\beta$'s in the ith precinct. If modeling fractions and
proceeding in a Bayesian fashion, a researcher might put a prior on the
$\uBi$'s with parameter ${\underaccent{\sim}{\zeta}}$, in which case one
representation of this class of models is as follows:
\begin{equation}\label{e:eq1}
p({\underaccent{\sim}{\zeta}} | \Ncolbold, \Nrowbold)  \propto
p({\underaccent{\sim}{\zeta}}) \prod_{\mathrm{i} = 1}^{\mathrm{I}} \biggl[ \int
p(\uNicol
| \uBi, \uNirow) \times p(\uBi| {\underaccent{\sim}{\zeta}})\, d\uBi\biggr].
\end{equation}
For example, in the simplest version of the linear model,
$p(\uBi| {\underaccent{\sim}{\zeta}})$ can be conceptualized as a multivariate
normal with mean vector ${\underaccent{\sim}{\beta}}$ and null variance. In
\citet{Rosen2001}, $p({\underaccent{\sim}{\zeta}})$ is a set of mutually
independent univariate gamma distributions,
$p(\uBi|{\underaccent{\sim}{\zeta}})$ a product Dirichlet, and $p(\uNicol|
\uBi, \uNirow)$ a multinomial parameterized by a mixture of $\beta$'s
and the fractions produced when $\uNirow$ is divided by its sum.
Particularly important is the fact that in proportionality (\ref{e:eq1}),
because there is no distribution posited for the unobserved internal
cell counts, there is no summation needed to eliminate them. [Note that
throughout this paper, including in proportionality (\ref{e:eq1}), we have
written the models we fit in terms of posterior distributions for the
hyperparameters. We have done so because, as we will explain, interest
sometimes centers on these population-level hyperparameters. As a
practical matter, the Markov chain Monte Carlo (MCMC) algorithms used
to fit these models typically work on the full joint posterior of all
model parameters. For more detail on model fitting, see Appendix A.2
of \citet{GreinerQuinn2009}.]

In contrast, consider a class of techniques that models the unobserved
internal cell counts. A researcher proceeding in a manner analogous to
proportionality (\ref{e:eq1}) might specify a distribution for each precinct's
internal cell counts given some precinct-level intermediate parameters
(call these intermediate parameters ${\underaccent{\sim}{\Upsilon}}_i$), might
specify a prior on the ${\underaccent{\sim}{\Upsilon}}_i$'s (call the parameters
in this prior ${\underaccent{\sim}{\Xi}}$), and might sum out the unobserved
internal cell counts. Thus, the proportionality corresponding to
(\ref{e:eq1}), above, is
\begin{eqnarray}\label{e:eq2}
&&p({\underaccent{\sim}{\Xi}} | \Ncolbold, \Nrowbold) \propto
p({\underaccent{\sim}{\Xi}}) \prod_{{\mathrm{i}} = 1}^{\mathrm{I}} \biggl[
\int\sum_{\Nimiss} p(\uNicol| \Nicomp)\nonumber\\[-8pt]\\[-8pt]
&&\hspace*{132pt}{} \times\
p(\Nicomp|{\underaccent{\sim}{\Upsilon}}_{\mathrm{i}}, \uNirow)
\times p({\underaccent{\sim}{\Upsilon}}_{\mathrm{i}}|\Xi)\,
d{\underaccent{\sim}{\Upsilon}}_{\mathrm{i}} \biggr]\nonumber.
\end{eqnarray}


$p(\uNicol| \Nicomp)$ appears to make the relationship
between the left- and right-hand sides of the $\propto$ symbol more
transparent; in fact, $\Nicomp$ determines $\uNicol$, rendering
$p(\uNicol| \Nicomp)$ degenerate. Note in this formulation there is
an explicit model for the internal cell counts
[$p(\Nicomp|\Upsilon_{\mathrm{i}}, \uNirow)$], which in turn requires a
summation over $\Nimiss$ to produce the observed-data likelihood. But
the distribution of $\Nimiss$ is complicated; the permissible support
of each element of $\Nimiss$ depends on the value of the other
elements. Further, in voting applications, the number of voters
involved is typically large enough to render infeasible full
computation of the posterior probabilities associated with every
permissible count.

Thus, proportionalities (\ref{e:eq1}) and (\ref{e:eq2}) make explicit the benefits
of each approach. By avoiding the need for a summation over a
complicated discrete distribution, proportionality (\ref{e:eq1}) makes fitting
easier. This benefit should not be understated. As we will discuss
below, the lack of information in ecological data can make model
fitting, even via MCMC, slow and cumbersome. The model we advocate
requires drawing from two multivariate distributions (one for the
internal cell counts, one for $\Upsilon_{\mathrm{i}}$) for each precinct for
each of a minimum of several hundred thousand iterations of an overall
Gibbs sampler. In contrast, the proposal in \citet{Rosen2001}, for
example, requires only one draw per precinct from a more standard
distribution, resulting in substantially less time to analyze a data
set.

The speed gain has trade-offs. For the purposes of this paper, the
primary down side is the lack of an easily conceptualized way of
incorporating individual-level information into the model due to the
lack of an explicit distribution $p(\Nicomp| {\underaccent{\sim}{\Upsilon}
}_{\mathrm{i}}, \uNirow)$. In contrast to proportionality (\ref{e:eq1}), proportionality
(\ref{e:eq2}) can be modified in a simple way to incorporate data from a
sample, as follows:
\begin{eqnarray}\label{e:EIwithSample}
&&p({\underaccent{\sim}{\Xi}} | \Kbold, \Ncolbold, \Nrowbold) \propto
p(\Xi)
\prod_{\mathrm{i} = 1}^{\mathrm{I}} \biggl[
\int\sum_{\Nimiss} p(\Kibold| \Nicomp)^{(\mathrm{i}
\in\mathrm{S})}\nonumber \\
&&\hspace*{145pt}{}\times p(\uNicol| \Nicomp)
\times p(\Nicomp|{\underaccent{\sim}{\Upsilon}}_{\mathrm{i}}, \uNirow) \\
&&\hspace*{259pt}{} \times
p({\underaccent{\sim}{\Upsilon}}_{\mathrm{i}}|{\underaccent{\sim}{\Xi}})
\,d{\underaccent{\sim}{\Upsilon}}_{\mathrm{i}} \biggr].\nonumber
\end{eqnarray}
Additional costs, discussed in \citet{GreinerQuinn2009}, to
the approach in proportionality (\ref{e:eq1}) are the difficulty in articulating
an individual-level (voter) conceptualization of the underlying
data-generating process [assuming one is desirable, see
\citet{King1997} for a different view] and the fact that most such
models weight contingency tables equally regardless of size.

Proportionality (\ref{e:EIwithSample}) further demonstrates that this
formulation allows for any within-contingency-table sampling scheme to
be implemented, so long as one can write down $p(\Kibold| \Nicomp)$.
Note, however, that the exchangeability assumption (reflected in the
product over i) prevents incorporation of contingency-table-level
sample weights into the likelihood. In other words, proportionality
(\ref{e:EIwithSample}) does not take into account whether the contingency
tables in S are selected via simple random sampling, sampling in
proportion to size, etc. As we explain below, this fact can be a
strength or a weakness, but whichever it is, it does not mean that all
contingency-table sampling schemes are equally beneficial.

Finally, proportionality (\ref{e:EIwithSample}) demonstrates that a variety of
choices of likelihoods, priors and hyperpriors for count models are
available. We next discuss our choices.

\section{Our proposal}\label{s2}

In the line of proportionality (\ref{e:EIwithSample}), our proposal consists
of the following. For $\Nicomp|{\underaccent{\sim}{\Upsilon}}_{\mathrm{i}}, \uNirow$, we assume that the counts in each contingency table
row follow an (independent) multinomial distribution with count
parameter N$_{r_i}$ and probability parameter $\uthri$. We choose the
multinomial because it corresponds to an individual-level account of
voting behavior (each potential voter of race r in precinct i
independently behaves according to the same vector $\uthri$) and
because once one conditions on the row totals (as is customary in
voting applications), few other tractable multivariate count
distributions are available.

For ${\underaccent{\sim}{\Upsilon}}_i|{\underaccent{\sim}{\Xi}}$, we apply a
multidimensional additive logistic transformation
[\citet{Aitchison2003}] to each row's $\uthri$, resulting in R vectors
of dimension (C-1), which we stack to form a single vector $\uomi$ of
dimension R${}\times{}$(C$-$1) for each precinct. We then assume $\uomi
\stackrel{\mathrm{i.i.d.}}{\sim} N(\umu, \SI)$. We prefer the multidimensional
additive logistic to, say, a Dirichlet or a different transformation
because of the additive logistic's greater flexibility relative to the
Dirichlet [\citet{Aitchison2003}] and because of the intuitive choice
of a ``reference category'' in voting applications, namely, the Abstain
column. The stacking of the transformed $\uthri$'s into a single
vector allows for exploration of within- and between-row relationships;
as we demonstrate in our application (see Figure
\ref{fig:AsianCorrPlot}), capacity to model between-row relationships
can be important to inference.

For the hyperprior [$p({\underaccent{\sim}{\Xi}})$], we use semi-conjugate
multivariate normal and inverse Wishart forms, specifically $\umu\sim
N(\umuz, \kazbold)$ and $\operatorname{\bolds{\Sigma}\sim Inv\mbox{--}Wish}_{\nuz}(\PSz)$. We do so both
for computational convenience and because, after extensive simulations,
we have found these distributions rich enough to express most
reasonable prior beliefs regarding the content of the contingency
tables.

For $p(\Kibold| \Nicomp)$, we assume a simple random sample, out of
necessity. Several recent papers [e.g.,
\citet{GlynnEtAl2008}, \citet{HaneuseWakefield2008} and
\citet{GlynnEtAl2009}] have discussed optimal within-contingency-table
sampling designs, with the optimal scheme varying according to the
process assumed to generate the data and to whether one of the rows or
columns corresponds to a relatively rare event (often true in
epidemiology applications). All of these schemes depend on the
assumption that the researcher can observe some characteristic of an
individual unit before deciding whether to include it in the sample.
This is not always possible in exit polls because voters exit polling
locations rapidly and, for this reason, exit polls are often interval
samples, with the assumption that the interval produces a random sample
made plausible by keeping the interval at reasonable length.

If the exit poll constitutes a simple random sample in each i $\in$ S,
we can work with the R${}\times{}$C-dimension vector $\uKi$ formed by summing
$\Kibold$'s columns; this results in a vector of counts of the number
of sampled potential voters in each contingency table cell, with the
contingency table vectorized row major. Denote the elements of $\uKi$
as $\Kirc$, $\Ki= \sum_{\mathrm{r, c}} \Kirc$ and for each i $\in$ S, recall
$\Mirc= \Nirc- \Kirc$. Accordingly, the probability of observing a
particular vector $\uKi$ is the familiar $\frac{{\prod}_{\mathrm{r,c}}{{\Mirc+ \Kirc}\choose{\Kirc}}}{{{\Ni}\choose{\Ki}}}$  [see \citet{McCullaghNelder1989}].

Upon discarding terms for i $\in$ S that do not involve unobserved
quantities, combining terms, canceling and including the Jacobian of
the transformation from $\theta$ to $\omega$ space, our proposal has
the following observed-data posterior:
\begin{eqnarray}\label{e:eq4}
&&p(\umu, \SI| \Kbold, \Ncolbold, \Nrowbold))\nonumber\\
&&\qquad \propto \mathrm{N}(\umu|\umuz, \kazbold) \times \operatorname{Inv\mbox{--}Wish}_{\nuz}(\SI| \PSz) \nonumber\\[-8pt]\\[-8pt]
&&\qquad\quad{} \times \prod_{\mathrm{i}} \biggl[ \int\biggl(
\sum_{\Mimiss} \biggl( \prod_{\mathrm{r,c}}
\frac{\thrci^{\Mirc}}{\Mirc!} \biggr)\biggr)^{{\mathrm{i}} \in
{\mathrm{S}}} \biggl( \sum_{\Nimiss} \biggl( \prod_{\mathrm{r,c}}
\frac{\thrci^{\Nirc}}{\Nirc!} \biggr) \biggr)^{{\mathrm{i}} \notin
{\mathrm{S}}} \nonumber\\
&&\qquad\quad\hspace*{36pt}{}\times \biggl( |\Sigma|^{-1/2}
\exp\biggl\{ -\frac{1}{2} (\uomi- \umu)^T \Sigma^{-1} (\uomi-
\umu)
\biggr\} \biggr)\, d\uthi\biggr].\nonumber
\end{eqnarray}
 Proportionality (\ref{e:eq4}) can be understood as follows:
the first line represents the hyperprior. The second and third lines
correspond to the multinomial assumptions for the internal cell counts,
with the second line demonstrating one of the contributions that the
survey makes to the information in the posterior. As $\Kirc$ gets
large, $\Mirc= \Nirc- \Kirc$ decreases, reducing the uncertainty in
the exponent of the numerator of $\frac{\thrci^{\Mirc+ \Kirc
-1}}{\Mirc!}$ and driving the denominator to 1. If $\Ki= \Ni$
(meaning that all voters in precinct i were sampled), then this portion
of the posterior corresponds to the nonconstant portion of the
likelihood of the probability vector of a multinomial distribution.
The fourth line is the multivariate normal. Note that a fair amount of
structure is contained within the summations over $M_{\mathrm{miss}_{\mathrm{i}}}$ and $N_{\mathrm{miss}_{\mathrm{i}}}$ as well as the integral over $\uthi$.
In each precinct i, the missing internal cell counts must sum to their
row and column totals, and each contingency table row's $\theta$'s must
stay within a simplex. A more complex version of proportionality
(\ref{e:eq4}), which demonstrates more explicitly the constraints
involved, appears in \citet{suppA}.

In many voting applications, particularly in redistricting, quantities
represented above by Greek letters are of limited interest. Instead,
interest lies in functions of the counts produced upon summation of the
contingency tables over i. These functions include $\LArc=\frac{\sum_{\mathrm{i}} \Nirc}{\sum_{\mathrm{i}} ({\mathrm{N}_{r_\mathrm{i}}}- N_{\mathrm{rA}_{\mathrm{i}}})}$,
$\GAr= \frac{\sum_{\mathrm{i}} (\mathrm{N}_{r_\mathrm{i}}- N_{\mathrm{rA}_{\mathrm{i}}})}{\sum
_{\mathrm{i}}
(\Ni- \NiA)}$, and $\TOrc= \frac{\sum_{\mathrm{i}} ({\mathrm{N}_{r_\mathrm{i}}}-
N_{\mathrm{rA}_{\mathrm{i}}})}{\sum_{\mathrm{i}} {\mathrm{N}_{r_\mathrm{i}}}}$ representing, respectively, the
fraction of
actual (as opposed to potential) voters of race r supporting candidate
c, the fraction of actual voters who are of race r, and the turnout of
race r's potential voters. The interest in these (and other) functions
of the internal cell counts leads us to fit our proposal via a
three-part Gibbs sampler; details appear in \citet{suppA}.

Speed is a serious concern here. \citet{suppA} have some details, but
depending on the constraints imposed by the bounds, ecological data can
have little information in them, resulting in slow mixing. At present,
after experimenting with several choices of proposal distributions [see
\citet{MetropolisEtAl1953} and \citet{TannerWong1987}] and fitting
algorithms, our software run on a reasonable laptop can ordinarily
analyze a data set of the approximate size of a typical United States
congressional district in a few hours. As of now, then, analyzing
multiple data sets in a short period of time, a feature of some modern
United States voting rights litigation, may require special
computational tools. We continue to work to address this situation.

\section{A comparison of estimators}\label{s3}

We present the results of simulation studies primarily addressing two
broad questions. First, in the R${}\times{}$C context, what is the
relative performance of an ecological model alone, a survey estimator
alone and our hybrid technique? In particular, we are interested in
the relative performance of these three classes of estimators: (i) in
the presence or absence of aggregation bias, and (ii) when contingency
tables have relatively even distribution of counts among rows versus a
moderate tendency for counts to be concentrated in one or another row.
Note that if counts in contingency tables tend to be distributed
relatively evenly among the rows, the bounds [\citet{DuncanDavis1953}]
constrain the posterior less. In voting parlance, segregated housing
patterns tend to lead to better performance of an ecological model.

Our second question of interest is whether the method of selecting the
contingency tables (precincts) for inclusion in the sample S affects
estimation. The advantages of probability weighting according to some
observed criteria, such as size, are well understood in the survey
literature. In the context of ecological data, however, we are
interested in whether any benefits accrue to weighting contingency
tables according to whether their bounds were likely to constrain,
that is, whether a particular table's counts were mostly in
one row. In voting parlance, is there an advantage to weighting
racially uniform precincts differently from racially mixed precincts?

\subsection{Simulation methods}\label{s31}

We simulated blocks of 100 voting jurisdictions, producing data sets
that generally resembled a United States congressional district in
which a court might look for racial bloc voting. We assumed three
racial groups (black, white, Hispanic) and two candidates (Democrat,
Republican), producing precinct-level tables as per Table
\ref{tbl:3.3.counts.full}. For each jurisdiction, we applied seven
estimation techniques: an ecological model alone; three two-stage
sampling estimators in which sampled precincts were selected using
different weighting schemes, after which a simple random sample was
taken of potential voters within each precinct; and three hybrid
estimators, in which the ecological model was combined with the data
from each of the two-stage samples. With respect to the three survey
samples, the first (``Sampling Scheme 1'') assigned much heavier
weights to racially integrated precincts, the second (``Sampling Scheme
2'') applied moderately greater weights to racially integrated
precincts, and the third (``Sampling Scheme 3'') applied much heavier
weights to racially uniform precincts. Population fractions, turnout
levels and party preferences of blacks, whites and Hispanics were set
at levels approximating behavior we have observed in United States
congressional districts.


We present the results for six simulated blocks of jurisdictions:
integrated (less integrated) without aggregation bias; integrated (less
integrated) with aggregation bias; and integrated (less integrated)
with severe aggregation bias. To induce aggregation bias, we turned
the top-level (normal distribution) location parameters for whites
($\mu_{wD}$ and $\mu_{wR}$) into linear functions of the fraction
Hispanic, $\Xih= \frac{\Nih}{\Ni}$. After speaking to a few persons
knowledgeable in the field of voting rights and racial bloc voting
regarding what might be realistic, we chose figures for aggregation
bias that, in expectation, would induce white voters in an 20\%
Hispanic precinct to vote approximately 70\% for the Republican, while
white voters in an 80\% Hispanic precinct would vote 55\% for the
Republican. The corresponding figures for the severe aggregation bias
(an approximately 90\% to 30\% swing in white Republican support) were
designed to be unrealistically harsh and to test the outer limits of
the method. We present here the results for the quantity $\LAhD$
because, in voting applications, it is often difficult to estimate and
particulary vulnerable to aggregation bias, with both of these factors
due to Hispanics' lower turnout rates and greater tendency (relative to
blacks) to vote in nonuniform patterns.



\begin{figure}

\includegraphics{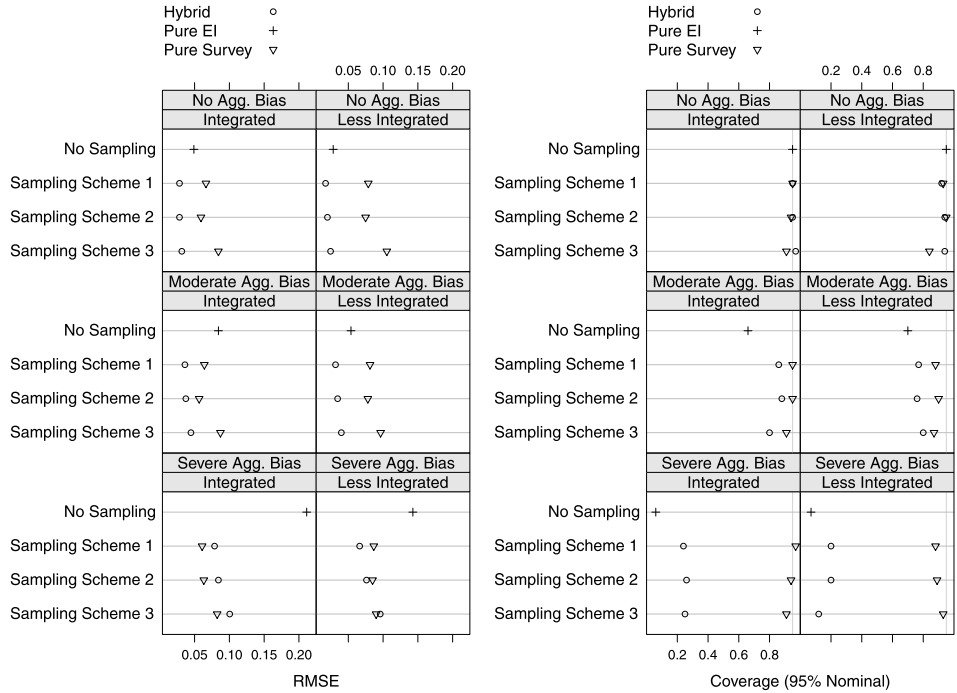}

\caption{Summary of results from simulations. The left
panels display RMSE, while the right panels display the
coverage of nominal 95\% credible intervals. Sampling
Scheme 1 heavily overweights racially mixed precincts,
Sampling Scheme 2 mildly overweights racially mixed
precincts, and Sampling Scheme 3 heavily overweights
racially uniform precincts. Note that ``Integrated''
data sets have less information in the bounds. The results show that the
hybrid estimator generally outperforms the pure survey and
pure ecological inference estimators, and offers
substantial RMSE reductions in many circumstances. In the
absence of severe aggregation bias, the hybrid estimator's
coverage is typically less than but comparable to that of
the pure survey estimator. } \label{fig:RMSE}
\end{figure}

When comparing estimators, we proceed on several of the usual fronts,
examining coverage of 95\% intervals, 95\% interval length and root
mean squared error (``RMSE''). In addition, because we apply the same
seven estimation techniques to each simulated data set, we examine how
often estimators outperform one another in each simulation block in
terms of squared error by calculating a binomial \textit{p}-value under a null
hypothesis of that the two estimators compared are the same. When we
report a \textit{p}-value, we mean this value unless we state otherwise.

Additional details of our simulations appear in \citet{suppA}.

\subsection{Simulation results}\label{s32}

Basic results are summarized in Figure \ref{fig:RMSE}. We draw the
following conclusions. First, hybrid estimators trounce the pure
ecological inference estimator under all circumstances. While we do
not find this result surprising in the abstract, the magnitude of the
improvement is worthy of note. In the absence of aggregation bias, the
hybrid estimators offer greater precision, producing posterior
intervals that are narrower but that still provide stochastically
nominal coverage. The best-performing hybrid (Sampling Scheme 1)
results in a reduction of posterior interval length of approximately
30--50\%, depending on the level of integration in housing patterns.
With aggregation bias, the hybrid raises the coverage of the 95\%
intervals from poor (roughly 0.68) to a level that, while less than
nominal, might approach tolerability (roughly 0.85). Meanwhile, the
RMSE reductions are on the order of 30--60\%. With severe aggregation
bias, any estimator that uses the ecological data fails to achieve
nominal coverage. Nevertheless, all hybrids substantially outperform
the ecological estimator alone. The reduction in RMSE, on the order of
55\%, is substantial, with this result stemming from both a noticeable
decrease in bias and a noticeable increase in precision. In comparing
any hybrid to the ecological inference estimator, all \textit{p}-values from our
simulations are 0. From this, we provide the following recommendation:
always include the survey.

Second, comparing hybrids to one another, there are advantages to
avoiding a sampling scheme that oversamples contingency tables in which
one row dominates, that is, racially homogenous precincts.
Without aggregation bias, the difference between the hybrid that
oversamples racially homogenous precincts (Sampling Scheme 3) versus
the other two (Sampling Schemes 1 and 2, which overweight racially
mixed precincts) is noticeable but modest; the latter offer 10--20\%
reductions in 95\% interval length (all \textit{p}-values less than 0.01). With
aggregation bias or severe aggregation bias, the improvement is larger.
The lack of nominal coverage makes 95\% interval length less
informative. But regarding RMSE, Sampling Scheme 1, which oversamples
racially mixed precincts, achieves 20--30\% reduction as compared to
Sampling Scheme 3, which oversamples racially uniform precincts (all
\textit{p}-values are 0).

The most difficult comparison is the hybrid estimators versus the pure
survey estimators. In the absence of aggregation bias, the conclusion
is simple, with any hybrid estimator constituting an enormous
improvement. The greater precision of the hybrid estimators is
reflected in both the length of the 95\% intervals, which can be as
much as 70\% narrower, as well as RMSE comparisons. Any hybrid
outperforms any pure survey estimator (all \textit{p}-values are 0).

With aggregation bias, we again recommend the hybrid over the pure
survey estimator, but we do so more cautiously. Although the pure
survey estimators' intervals come closer than the hybrids to achieving
nominal coverage, the coverage gains are modest (around 7\%).
Meanwhile, the RMSE gains from the hybrids, on the order of 35--60\%,
are substantial. On average, the bias of the hybrid estimates is
modest, roughly two or three percentage points (i.e., a point
estimate of 0.53 when the truth is 0.51). Thus, even in the presence of
aggregation bias, the hybrids offer substantial benefits over the pure
survey estimators.

In the presence of severe aggregation bias, the results are mixed.
With integrated housing patterns and in the presence of severe
aggregation bias, the combination of bias and lack of bounding
information renders the pure survey estimators superior, with hybrid
RMSEs approximately 10--20\% larger than their pure survey counterparts.
With severe aggregation bias and with less integrated housing patterns,
interval coverage for both types of estimators was less than nominal
(and worse for the hybrids). With respect to RMSE, however, on
average, the hybrids usually outperform their specific pure survey
counterparts, and the reductions are on the order of 10\% to as high as
25\%. Average does not mean always, however. And on a
simulation-by-simulation basis, the comparison of some pure survey
estimators to the hybrids results in \textit{p}-values near 0 in favor of the
pure survey estimators (recall that our \textit{p}-values represent which method
prevails simulation-by-simulation, a 0--1 outcome). The reason for this
is that the higher variances associated with the pure survey estimators
mean that when these estimators miss the target, they can miss badly,
raising the RMSE, which as a function of an average is sensitive to
large misses. In the presence of contextual effects, the
lower-variance hybrid estimators reduce the risk of a point estimate
that is badly wide of the mark, at the cost of some bias.

\subsection{Simulation conclusions}\label{s33}

Thus, as between hybrid versus survey estimators, which estimator
should a researcher prefer? In our view, the answer depends primarily
on three factors: the extent to which contingency tables tend to be
dominated by one row (i.e., the extent of racial segregation in
housing patterns), the magnitude of aggregation bias in the data, and
whether the ultimate user cares more about an accurate point estimate
or a valid interval. The first factor is observable. The second is
not observable, and it may or may not be that, in some instances, a
researcher or expert witness will have some information about
aggregation bias from external sources. Regarding the third, some
users pay attention primarily to point estimates. Courts, for example,
who in voting rights litigation may examine results from dozens of
elections, typically do not incorporate uncertainty estimates into
their opinions, despite exhortations from social scientists to the
contrary. Other users make what we suspect for statisticians is the
more traditional choice. In general, however, our recommendation is
that unless the researcher has reason to fear extremely strong
(``severe'' really means ``brutal'') aggregation bias, the hybrid
estimator is preferable.


\section{Boston area colleges exit poll}\label{s4}

Did Asian-American voters in the City of Boston support a Massachusetts
ballot initiative repealing criminal penalties for possession of small
amounts of marijuana? Were
support rates for the marijuana initiative different between Caucasian
versus Hispanic voters? To test the methods we propose, we conducted
an exit poll in the City of Boston on November 4, 2008. Because our
interest is in both the operational feasibility as well as the
comparative technical advantages or disadvantages of hybrid estimators,
we briefly describe the running of the poll and the necessary
preprocessing of the data before articulating required assumptions and
providing results. We demonstrate that the two questions articulated
above are difficult to answer with either the exit poll or the
ecological estimator standing alone, but that the hybrid permits
reasonable inferences as to both.

\subsection{Mechanics and initial results}\label{s41}

We recruited law, graduate and undergraduate students from 11 Boston
area colleges and universities to participate in an exit poll. Our
recruiting efforts yielded over 400 pollsters, which we organized into
teams captained by a law or graduate student. There were two election
day shifts lasting seven hours each, which covered the whole of the
election day. Captains attended one of several 90-minute training
sessions, while training for noncaptain pollsters lasted an hour. All
sessions were live and covered essential survey/exit polling
techniques. For example, pollsters were instructed to step away from
voters after making a successful approach and to request that voters
themselves place completed questionnaires in a visibly closed box [see
\citet{BishopFisher1995}]. Five specially trained, two-person roving
quality control teams circulated in cars, visiting each polling
location multiple times throughout election day and monitoring
compliance with the required techniques. We attempted to deploy
multilingual pollsters to locations in which a comparatively high
percentage of voters spoke languages other than English.

Pollsters approached every eighth voter but alternated between a
``voter choices'' questionnaire, which generated the data used in this
paper, and a ``voter experience'' form, which was used for other
purposes. Effectively, this meant a targeted $\frac{1}{16}$ sampling
interval for the race-and-voter-choices exit poll. Prior coordination
with the City of Boston Election Department, together with the absence
of a law in Massachusetts regulating exit polls, enabled pollsters to
stand immediately outside the exits to the buildings in which voting
occurred, and teams were large enough to cover all exits.

The poll covered 39 of Boston's 160-odd polling locations. 26 of the
39 locations were selected in a nonrandom manner due to the research
design associated with the voter experience questionnaire; the other 13
were randomly selected using inverted Herfindahl index weights that
resulted in a higher probability of selecting polling locations in
which several racial groups were represented [see \citet{suppA} for
details].

\begin{figure}[b]

\includegraphics{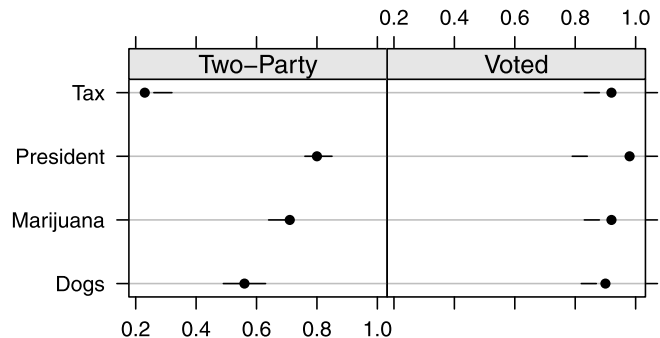}

\caption{Results of Boston Area Colleges Exit Poll (pure
survey estimators). ``Two-Party'' refers to the percentage of
actual voters voting for Obama (Presidential) or Yes (income
tax, marijuana and dog racing ballot initiatives), while
``Voted'' refers to the percentage of persons entering the
ballot who cast ballots in the relevant contest. True values
are represented by solid dots, 95\% confidence
intervals are represented by the dark lines. Two-party point estimates
are generally accurate, but nonvoting behavior is
overestimated.}\label{f:tab2}
\end{figure}

Overall, Boston Area Colleges Exit Poll pollsters approached
approximately 4300 voters with voter choice questionnaires and achieved
approximately a 57\% response rate. Voter choice data were collected
for United States president and for three Massachusetts ballot
initiatives, one repealing the state income tax, one eliminating
criminal penalties for possession of small amounts of marijuana, and
one banning gambling on dog racing. After multiply imputing for
nonresponse (see below), we applied a stratified (to reflect the
separate deterministic versus random precinct-selection schemes),
two-stage (cluster followed by simple random sample) estimator to the
results to check our predictions against the known truth. As Figure
\ref{f:tab2} demonstrates, we found that our projections closely
approximated the overall true two-party vote fractions, where
``two-party'' means the percentage of Obama supporters out of those who
voted for either Obama or McCain, or the percentage of Yes votes out of
those who voted Yes or No on the ballot initiatives. We did find,
however, a curious [see \citet{Silver1986}] tendency among poll
respondents to overreport nonvoting behavior, and the prior in our
multiple imputation algorithm may have exaggerated this aspect of the
data. For these reasons, we compare estimators for the marijuana
initiative, where our two-party projection was accurate, where
nonvoting overreport was comparatively low, and where the two-party
vote was reasonably close. Results for the dog racing ballot
initiative, which share these characteristics, were similar, and are
available from the authors.

\subsection{Data processing and critical assumptions}\label{s42}

We detail in this section the critical assumptions underlying our
various estimators. First, to account for nonresponse, we created 10
completed data sets via multiple imputation. The imputation model was a
loglinear model for categorical data as implemented in Joe Shafer's
\texttt{cat}
package.\footnote{\url{http://cran.r-project.org/web/packages/cat/index.html}.}
Computational challenges arose because of the fairly large number of
variables to impute and our desire to allow for more complicated
associations than would be possible under a multivariate normal model
or a 2-way loglinear model. To overcome these challenges, we made use
of a parametric bootstrap approach [\citet{HonKin09}] along with a
factorization of the full data distribution that allowed us to work
with the data in moderately-sized chunks.

Our procedure was the following. First, we created 10 bootstrap data
sets by sampling rows with replacement from the observed data matrix.
We partitioned the variables in each of these bootstrap data sets into
three sets---pollster-specific attributes, voter demographics and voter
choice variables. Then, for each of the bootstrap data sets, we imputed
pollster-specific attributes, voter demographics given the imputed
pollster attributes, and finally voter choice data given the imputed
voter demographics and a subset of the imputed pollster
characteristics.

Each imputation step worked as follows. Given a particular bootstrap
data set, we calculated the posterior mode of the cell probabilities
using the ECM algorithm. We then sampled the missing data from the
appropriate multinomial distribution with probabilities given by the
maximum a posteriori estimates. For the pollster-specific data (which
had very little missingness) and the voter demographic data we employed
a loglinear model with all 3-way interactions and a Dirichlet prior for
the cell probabilities with parameters all equal to 1.0001. For the
voter choice data (which had more missingness) we used a loglinear
model with all 2-way interactions and a Dirichlet prior on the cell
probabilities with parameters equal to 1.001.

The assumptions underlying the multiple imputations are the primary
ones needed to render the pure survey estimators discussed below valid.
Another assumption is that the interval sample produced a simple random
sample of voters in the in-sample precincts. We deem this assumption
plausible in light of the $\frac{1}{16}$ target interval. Overall, in
assessing these assumptions, for the two electoral contests presented
in this paper, we are encouraged by the exit poll's ability to project
closely the two-party vote and to approximate the amount of nonvoting
observed in the ballot initiatives.

For the hybrid and pure ecological estimators, the most important
assumption is lack of contextual effects. With respect to this data
set, however, the no-contextual-effect assumption is slightly stronger
for the data as used by the hybrid estimator than for the data as used
by the pure ecological counterpart because the hybrid operates on more
aggregated data, as follows. Exit polls survey voters by polling
location, not by precinct, and in the City of Boston, many polling
locations host voters of more than one precinct such that pollsters
standing outside of a polling location's building are unable to
distinguish voters from the various precincts housed there. Thus, the
data used by the hybrid estimator had to consist of figures at the
level of the polling location (for in-sample polling locations),
that is, further aggregated. The ecological estimator could
operate at the level of the individual precinct. Note that for the
data used by the hybrid, for out-of-sample polling locations, we used
precinct-level (as opposed to polling-location-level) figures. Note
also that, at least in Boston, precincts are never split into two or
more polling locations; that is, each precinct is contained wholly
within one polling location.

\begin{figure}

\includegraphics{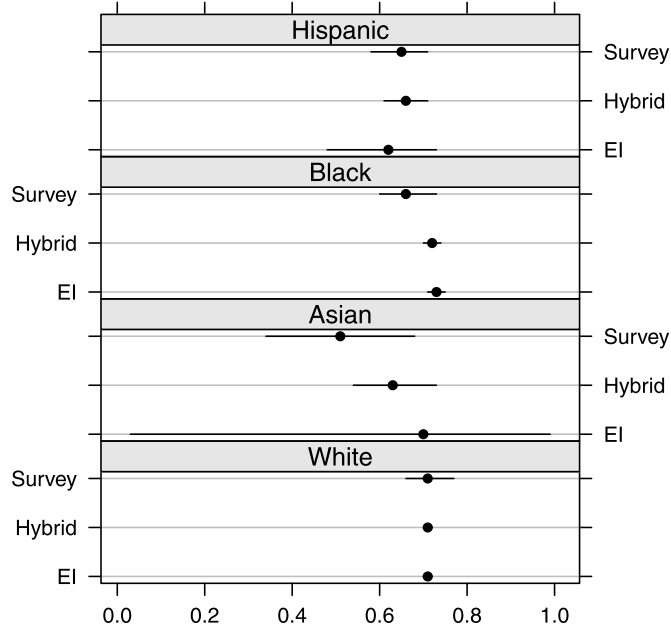}

\caption{Estimated fractions of support for the
marijuana decriminalization ballot
initiative among the four most numerous racial groups in the
City of Boston. Filled circles represent point estimates and
dark lines represent 95\% credible intervals. ``EI'' refers to
ecological inference estimator alone, ``Survey'' is the exit
poll alone, and ``Hybrid'' is the hybrid estimator. Survey and
Hybrid estimates come from multiple imputation. Only
the hybrid estimator offers enough precision in the Asian
category to allow substantive inference. The hybrid estimator also best
differentiates Hispanic from white preferences.}\label{f:mj-dog}
\end{figure}

According to figures based on Census 2000 and provided by the Boston
Redevelopment Authority, the City of Boston's voting age percentages by
race are as follows: 55\% white, 20\% black, 12\% Hispanic, 9\% Asian,
and the rest of ``other'' race.\footnote{Recalling that Census 2000
allowed respondents to mark more than one race box, these categories
are in fact shorthand for the following: ``Hispanic'' means Hispanic
(regardless of any other race box checked), ``Asian'' means
non-Hispanic any part Asian, ``black'' means non-Asian non-Hispanic any
part black, and ``white'' means anyone left who was not in the other
race category.} We investigated whether the various estimators under
consideration could say anything useful about Boston's four most
populous racial groups.

\subsection{Results of various estimators: Voting preferences by race}\label{s43}

Our results are encapsulated in Figure \ref{f:mj-dog}. We draw the following conclusions. First,
there is little evidence to contradict the critical no-aggregation-bias
assumption needed for the ecological and hybrid estimators. The point
estimates from the survey estimator generally align with those from the
other two. This fact does not provide total security, given the high
variance of the survey estimator, but total security is rarely
available when analyzing ecological data.

\begin{figure}

\includegraphics{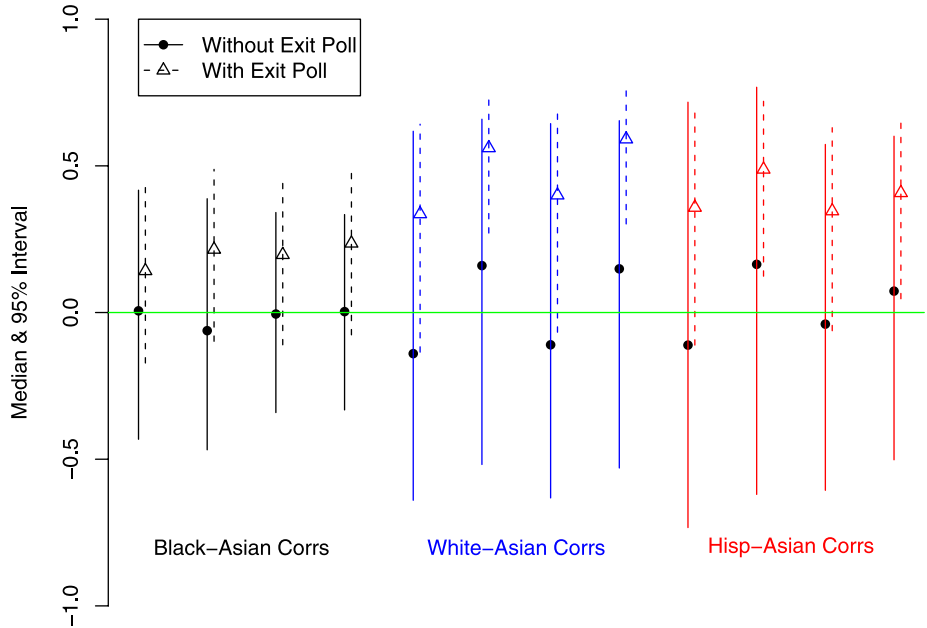}

\caption{Comparison of posterior distributions from
ecological inference, with and without exit poll, of
between-contingency-table-row correlations governing the
relationship of black, white and Hispanic voters to Asian
voters. (The with-exit-poll figures are averages of ten
multiple imputations.) The narrower posterior intervals,
and the greater density above zero, in the with-exit-poll
correlations suggest that the with-exit-poll model is
taking advantage of a tendency of various racial groups to
vote similarly within a precinct to provide better
estimates of Asian voting behavior. The without-exit-poll
model is unable to take advantage of this tendency.}\label{fig:AsianCorrPlot}
\end{figure}

Second, even after accounting for nonresponse via multiple imputation,
which necessarily involves higher variances than would be present for a
survey without nonresponse, the hybrid estimator provides substantial
variance reduction in a way that makes a substantive difference. For
example, in the marijuana ballot initiative, the 95\% interval for the
Asian support rate was (0.03, 0.99) for the pure ecological inference
estimator and was (0.34, 0.68) for the pure survey, but the hybrid
interval was (0.54, 0.73). Thus, only via the hybrid estimator would a
researcher or an expert witness be able to conclude that Asian voters
in the City of Boston supported the marijuana initiative. The same
phenomenon occurs in the Asian vote on the initiative to ban gambling
on greyhound racing (results not shown). Further, the pure survey and
the pure ecological estimators are less able to distinguish Hispanic
versus white preferences regarding the marijuana initiative. For the
hybrid estimator, in contrast, these 95\% confidence intervals
intersect by only a hair's breadth.

These results are substantively interesting in their own right, but we
are encouraged by the fact that the hybrid estimator appears to help
where help is most needed. The variance reduction available for the
estimates of Asian and Hispanic voting behavior is substantial. As the
two racial groups with the lowest VAP and lowest turnout, Hispanics and
Asians represent the most difficult challenge to inference about voting
behavior by race, and the performance of the hybrid estimator here is
encouraging.

A question arises: how could this happen? How could the combination
of information from a survey and from ecological data, neither of which
alone provides useful results, reduce variance enough to allow for
meaningful substantive inference? We offer the hypothesis that the
answer lies in the better estimation of parameters governing
between-contingency-table-row (as opposed to
within-contingency-table-row) relationships. An example to clarify
this distinction: a~within-contingency-table-row relationship would be
a tendency for precincts that have high counts of Asians voting
Democrat to also have high counts of Asians who abstain from voting. A
between-contingency-table-row relationship would be a tendency for
precincts that have high counts of Asians voting Democrat to also have
high counts of blacks who vote Democrat.

Several commentators [e.g., \citet{King1997}] have noted the
difficulty in estimating model parameters that govern behavior between
(as opposed to within) contingency table rows. We explored the
relative paucity of information about between-row relationships in
\citet{GreinerQuinn2009}. It appears, however, that individual-level
data can stabilize estimates of between-row parameters in an important
way. Recall that in our model, we stack the logistic-transformed
probability vectors from each contingency table's row multinomial to
form a single vector of dimension R${}\times{}$(C$-$1), which we then assume follows
a multivariate normal. Accordingly, the covariance matrix of this
normal ($\SI$) can be decomposed into block diagonal elements, which
govern within-contingency-table-row relationships, and block
off-diagonal elements, which govern between-contingency-table-row
relationships. As applied to the City of Boston, with black, white,
Hispanic and Asian racial groups, the matrix is as follows:
\begin{eqnarray*}
\SI=\left[
\matrix{
\SIb& \SIbw& \SIbh& \SIba\cr
\SIbw& \SIw& \SIwh& \SIwa\cr
\SIbh& \SIwh& \SIh& \SIha\cr
\SIba& \SIwa& \SIha& \SIa
}
\right].
\end{eqnarray*}
Note that each of $\SIba$, $\SIwa$ and $\SIha$ is of
dimension 2 $\times$ 2, and because each is off the main diagonal, each
has four correlations within it.

It appears that the introduction of individual-level information allows
estimation of Asian voting behavior to borrow strength from estimates
of white, black and Hispanic voting behavior by way of better and more
precise estimation of the correlations in $\SIba$, $\SIwa$ and $\SIha$.
Figure \ref{fig:AsianCorrPlot} compares the posterior intervals of
these correlations in the marijuana ballot initiative in the pure EI
model versus the hybrid. The narrower intervals of the correlations
from the hybrid, together with the fact that most of the distributions
from the hybrid have most of their mass above 0, appear to enable
better modeling of between-contingency-table-row relationships. In
other words, the correlations represented suggest that within a
precinct, Asian voting behavior is similar to that of other racial
groups, particularly that of whites. We hypothesize that this
similarity, together with the substantial information about white
voting behavior (from the bounds), in turn allows non-Asian voting
behavior to inform estimation of Asian preferences. If we are right,
this fact highlights the importance of using a model flexible enough to
allow estimation of between-contingency-table-row relationships,
something few other R${}\times{}$C models do.

\section{Conclusion}\label{s5}

In this paper we have proposed a hybrid count ecological inference
model capable of handling data sets with contingency tables of any size
and shape. We have briefly explored the benefits of count versus
fraction models in the R${}\times{}$C context as well as the implications
of different contingency-table-level sampling schemes. We have met the
challenge of operationalizing the use of our hybrid to voting data by
conducting an exit poll in the City of Boston, and in doing so have
confronted a difficult scenario for a hybrid estimator because of (i)~the impossibility of using optimal within-table sampling schemes, (ii)
the problem of nonresponse, (iii) the additional level of aggregation
occurring when more than one precinct share the same polling location,
and (iv) the desire to estimate behavior of groups with low VAP and
turnout. Our operationalization demonstrates that the hybrid model
offers benefits to those who seek inferences regarding racial voting
patterns.

\section*{Acknowledgments} The authors extend great
thanks to Jayanta Sircar at the Harvard School of Engineering for
allowing us to use the Harvard Crimson Grid gratis, and to Robert
Parrot for his assistance in teaching us how to use the grid; to
Gary King, Lewis Kaplow, Jeffrey Lewis, Andrew Martin and Kathy
Spier for helpful comments; and to the JEHT Foundation, the
Rappaport Institute for Greater Boston, the Boston Foundation, the
Sheldon Seevak Fund and the Harvard University Office of the
Provost for financial support. In addition, this paper benefited
greatly from feedback received during presentations at the Applied
Statistics Workshop at the Institute for Quantitative Social
Science at Harvard University; at the Law, Economics, and
Organization seminar at Harvard Law School; and at the 2009
Conference on Empirical Legal Studies.

%
\begin{supplement}
\sname{Supplement A}
  \stitle{Supplement to ``Exit polling and racial bloc voting:
    Combining individual-level and R${}\bolds{\times}{}$C ecological data''\\}
  \slink[doi]{10.1214/10-AOAS353SUPPA}
  \slink[url]{http://lib.stat.cmu.edu/aoas/353/SupplementA.gz}
  \sdatatype{.gz}
  \sdescription{This supplement describes the algorithms used to fit
    the models described in ``Exit polling and racial bloc voting:
    Combining individual-level and R${}{\times}{}$C ecological data.''}
\end{supplement}

\begin{supplement}
\sname{Supplement B}
  \stitle{Replication materials for ``Exit polling and racial bloc
    Voting: Combining individual-level and R${}\bolds{\times}{}$C ecological
    data''\\} \slink[doi]{10.1214/10-AOAS353SUPPB}
  \slink[url]{http://lib.stat.cmu.edu/aoas/353/SupplementB.tar.gz}
  \sdatatype{.gz}
  \sdescription{This supplement provides data and computer code that
    can be used to replicate the results in ``Exit polling and racial
    bloc voting: Combining individual-level and R${}\times{}$C
    ecological data.''}
\end{supplement}

\printaddresses

\end{document}